\documentclass[conference]{IEEEtran}
\IEEEoverridecommandlockouts
\usepackage{cite}
\usepackage{amsmath,amssymb,amsfonts}
\usepackage{algorithm,algorithmic}
\usepackage{xcolor}
\usepackage{soul}
\usepackage{tikz}
\usetikzlibrary{positioning, arrows.meta, shapes, calc,fit}
\usepackage{graphicx}
\usepackage{textcomp}
\usepackage{xcolor}
\def\BibTeX{{\rm B\kern-.05em{\sc i\kern-.025em b}\kern-.08em
    T\kern-.1667em\lower.7ex\hbox{E}\kern-.125emX}}
\begin{document}

\title{SliceFed: Federated Constrained Multi-Agent 
DRL for Dynamic Spectrum Slicing in 6G
}

\author{\IEEEauthorblockN{Hossein Mohammadi\IEEEauthorrefmark{1}, Seyed Bagher Hashemi Natanzi\IEEEauthorrefmark{2}, Ramak Nassiri\IEEEauthorrefmark{1}, Jamshid Hassanpour\IEEEauthorrefmark{3}, Bo Tang\IEEEauthorrefmark{2}, \\Vuk Marojevic\IEEEauthorrefmark{1}}
\IEEEauthorblockA{{\IEEEauthorrefmark{1}Department of Electrical and Computer Engineering,} 
{Mississippi State University,}
Starkville, MS, USA \\
\IEEEauthorrefmark{2}Department of Electrical and Computer Engineering, Worcester Polytechnic Institute, Worcester, MA, USA \\
\IEEEauthorrefmark{3}Department of Electrical and Computer Engineering, Georgia Institute of Technology, Atlanta, GA, USA\\
\{hm1125, rn533, vm602\}@msstate.edu, \{snatanzi, btang1\}@wpi.edu, 
\{jamshid\}@gatech.edu\\
}
}

\maketitle

\begin{abstract}
Dynamic spectrum slicing is a critical enabler for 6G Radio Access Networks (RANs), allowing the coexistence of heterogeneous services. 
However, optimizing resource allocation in dense, interference-limited deployments remains challenging due to non-stationary channel dynamics, strict Quality-of-Service (QoS) requirements, and the need for data privacy. In this paper, we propose SliceFed, a novel Federated Constrained Multi-Agent Deep Reinforcement Learning (F-MADRL) framework. SliceFed formulates the slicing problem as a Constrained Markov Decision Process (CMDP) where autonomous gNB agents maximize spectral efficiency while explicitly satisfying inter-cell interference budgets and hard ultra-reliable low-latency communication (URLLC) latency deadlines. We employ a Lagrangian primal-dual approach integrated with Proximal Policy Optimization (PPO) to enforce constraints, while Federated Averaging enables collaborative learning without exchanging raw local data. Extensive simulations in a dense multi-cell environment demonstrate that SliceFed converges to a stable, safety-aware policy. Unlike heuristic and unconstrained baselines, SliceFed achieves nearly 100\% satisfaction of 1~ms URLLC latency deadlines and exhibits superior robustness to traffic load variations, verifying its potential for reliable and scalable 6G spectrum management.
\end{abstract}

\begin{IEEEkeywords}
6G, Deep Reinforcement Learning, Federated Learning, Network Slicing, URLLC. 
\end{IEEEkeywords}

\section{Introduction}
\label{sec:introduction}

Dynamic and efficient radio spectrum utilization is a fundamental challenge for next-generation wireless systems. The proliferation of 5G deployments, massive IoT connectivity, and the coexistence of active networks with incumbent and passive sensing systems result in highly non-stationary and spatially heterogeneous interference conditions. Traditional rule-based sensing and thresholding methods, which typically assume stable or slowly varying spectral environments, struggle under bursty traffic, user mobility, and multipath propagation, leading to degraded reliability and responsiveness.

Reinforcement learning (RL) offers a principled framework for sequential decision-making under uncertainty by enabling an agent to learn optimal actions from interaction with the environment. RL based methods have demonstrated strong performance in dynamic communication settings, such as channel selection, distributed resource allocation, and link adaptation, particularly when the wireless channel or spectrum state evolves as a Markov decision process (MDP)~\cite{sutton1998reinforcement}, where future interference and channel states depend probabilistically on the current spectrum state and the agent’s actions. The ability of RL algorithms, such as Q-learning, double Q-learning, and deep Q-networks, to handle stochastic rewards and incomplete information makes them suitable for agile spectrum management tasks in congested, interference-limited environments. 

However, applying RL to spectrum sharing must confront two practical constraints: First, the agent must operate on measurements that exhibit rapid temporal variations caused by fading, shadowing, mobility, and device heterogeneity, as documented extensively in classical wireless channel models and empirical studies~\cite{goldsmith2005wireless}. Second, the learned policy must generalize to unseen interference patterns without relying on large, centrally curated datasets, which aligns with the broader trends of learning under limited observations and adapting online in complex environments.

This paper addresses these challenges by introducing a novel AI-driven spectrum management framework that unifies reinforcement learning with data-driven interference mitigation. The proposed method formulates spectrum access and interference avoidance as a constrained Markov decision process (CMDP), where the state is derived from calibrated spectral measurements and the agent selects continuous resource allocation actions under explicit interference and QoS constraints. Unlike conventional spectrum sharing approaches that rely on instantaneous power thresholds or receiver-side sensing alone, the proposed framework explicitly learns the \emph{temporal and cross-cell structure of interference leakage} and optimizes aggressor-side spectrum usage subject to hard constraint budgets. By integrating Lagrangian-based constrained learning with federated multi-agent coordination, the proposed approach enables adaptive, interference-aware spectrum access without predefined rules or centralized spectrum control.


The contributions of this paper are summarized as follows:
\begin{itemize}
    \item \textbf{SliceFed: A Federated Constrained Multi-Agent DRL Framework for Dynamic RAN Slicing.}
    We propose \emph{SliceFed}, a novel federated multi-agent deep reinforcement learning framework for dynamic radio access network slicing in dense B5G/6G deployments. SliceFed formulates per-gNB slicing decisions as local constrained Markov decision processes (CMDPs) with explicit inter-cell interference, QoS, and feasibility constraints, and solves them using a Lagrangian-based primal--dual PPO algorithm. A federated learning layer enables collaborative training across gNBs while preserving data locality and privacy, allowing decentralized execution with global coordination.

    \item \textbf{Constraint-Aware Spectrum and Resource Modeling.}
    We develop a rigorous system model that captures stochastic traffic dynamics, inter-cell interference leakage, and heterogeneous slice requirements, bridging wireless channel modeling with constrained reinforcement learning for spectrum-centric RAN slicing.

    \item \textbf{Stable and Low-Overhead Resource Adaptation.}
    The proposed framework incorporates reconfiguration cost awareness and dual-variable adaptation, enabling stable, low-variance resource allocation policies that avoid oscillatory behavior and excessive control signaling.

    \item \textbf{Comprehensive Evaluation and Benchmarking.}
    We evaluate SliceFed against representative baselines, including equal slicing, queue-based heuristics, and random allocation, demonstrating superior constraint satisfaction, stability, and robustness under increasing URLLC traffic loads.
\end{itemize}

Section~II reviews related work on reinforcement learning, constrained learning, and multi-agent approaches for network slicing. Section~III presents the system model and formulates the dynamic RAN slicing problem as a constrained Markov decision process (CMDP). Section~IV introduces the proposed Lagrangian-based constrained reinforcement learning framework. Section~V details the SliceFed architecture, including the federated multi-agent learning design and associated algorithms. Section~VI evaluates the performance of SliceFed through extensive simulations, and Section~VII concludes the paper.

\section{Literature Review}
\label{sec:literature_review}

Network slicing is a key enabler for supporting heterogeneous services with diverse quality-of-service (QoS) requirements in 5G and beyond wireless networks. Early optimization- and heuristic-based slicing approaches struggle to cope with highly dynamic traffic, stochastic channel conditions, and inter-slice coupling. As a result, reinforcement learning (RL), and in particular deep reinforcement learning (DRL), has been adopted as a data-driven alternative for adaptive resource allocation. Deep Q-learning and actor–critic frameworks have demonstrated notable gains over static baselines in terms of spectrum efficiency and quality-of-experience (QoE) under time-varying demand~\cite{4-8540003,6-9057705}. DRL-based slicing has also been studied in vehicular, smart-city, and edge-enabled scenarios, where latency-critical URLLC traffic must coexist with throughput-oriented services~\cite{7-9463417}. However, most existing DRL formulations treat slicing as an unconstrained MDP, relying on reward shaping and providing no formal guarantees on latency or reliability.

To explicitly incorporate QoS constraints, constrained reinforcement learning (CRL) and constrained MDP (CMDP) formulations have been introduced for network slicing. Xu \emph{et al.} proposed a CRL framework based on Lagrangian relaxation to enforce slice-level QoS constraints while optimizing long-term throughput~\cite{1-9333595}. The CLARA framework further formalized network slicing as a CMDP and demonstrated improved constraint satisfaction through budget-aware policy optimization~\cite{2-9671840}. Related efforts have extended constraint-aware DRL to end-to-end resource orchestration and industrial IoT scenarios, incorporating queue stability, latency, and reliability constraints~\cite{11-9651934,10-qi2022augmented}. While these works provide important theoretical foundations, they largely focus on single-cell or centralized control settings and do not explicitly capture inter-cell interference dynamics or distributed decision-making in dense cellular deployments.

More recently, multi-agent DRL has been investigated to enable scalable and coordinated slicing across multiple cells. Coordinated multi-agent DRL schemes have demonstrated improved delay performance and resource efficiency by mitigating inter-cell interference through limited information exchange or coordination among agents~\cite{8-9838518}. These approaches highlight the potential of distributed learning for large-scale radio access networks; however, most existing multi-agent solutions rely on unconstrained learning or heuristic coordination mechanisms, lacking a principled CMDP formulation with explicit constraint handling and dual-variable updates. As a result, strict QoS requirements—particularly for URLLC traffic—cannot be guaranteed under dynamic interference and traffic conditions.

In parallel, advanced DRL paradigms have been proposed to improve learning efficiency and robustness in network slicing. Distributional and generative approaches, such as GAN-powered DRL, have been shown to stabilize training under stochastic traffic demands~\cite{13-8931561}. Transfer learning and federated learning techniques have also been explored to accelerate convergence and enhance scalability by leveraging shared knowledge across slices or cells~\cite{9-9878155}. Nevertheless, the integration of federated multi-agent DRL with explicit CMDP-based constraint enforcement and spectrum-centric objectives—such as inter-cell interference control and URLLC reliability—remains largely unexplored in dense B5G/6G RANs. This gap motivates the proposed SliceFed framework.

More recently, service level agreement (SLA)-aware learning frameworks have gained attention, particularly in the context of O-RAN and industrial networks. Safe DRL methods aim to enforce reliability and latency constraints during both training and execution phases, addressing the risk of constraint violations inherent in exploration~\cite{12-nagib2025safeslice}. 
Researchers emphasize that future spectrum management and slicing solutions must explicitly integrate constraint handling, interference awareness, and distributed learning architectures to be viable in real-world deployments~\cite{14-10155733}. In contrast to existing work, this paper formulates dynamic RAN slicing as a CMDP with explicit spectrum, interference, and URLLC latency constraints and proposes a Lagrangian-based constrained RL framework suitable for distributed and federated implementations. By jointly addressing spectrum efficiency and hard QoS guarantees within a unified learning architecture, the proposed approach directly aligns with the spectrum management and advances the state of the art in AI-driven spectrum-aware network slicing.

\section{System Model and Problem Formulation}
\label{sec:system_model}


\begin{algorithm}[t]
    \caption{Local Primal--Dual Constrained RL at gNB $n$}
    \label{alg:local_primal_dual}
    \begin{algorithmic}[1]
        \STATE \textbf{Input:} Initial policy parameters $\theta_n$, critic parameters $\psi_n$, dual variables $\lambda_n = [\lambda_{1,n},\lambda_{2,n},\lambda_{3,n}]^\top \ge 0$, learning rates $\eta_\theta,\eta_\psi,\eta_\lambda$, discount factor $\gamma$
        \STATE Initialize replay buffer $\mathcal{D}_n \leftarrow \emptyset$
        \FOR{each local episode $e = 1,2,\ldots$}
            \STATE Observe initial state $s_n(0)$
            \FOR{each time step $t = 0,1,\ldots,T_e-1$}
                \STATE Select action $a_n(t) \sim \pi_{\theta_n}(\cdot \mid s_n(t))$
                \STATE Apply $a_n(t)$, observe next state $s_n(t\!+\!1)$, reward $r_n(t)$, and constraint signals $g_{i,n}(t)$, $i=1,2,3$
                \STATE Compute Lagrangian-adjusted reward:
                \[
                    \tilde{r}_n(t) = r_n(t) - \sum_{i=1}^3 \lambda_{i,n} g_{i,n}(t)
                \]
                \STATE Store transition $(s_n(t), a_n(t), \tilde{r}_n(t), s_n(t\!+\!1))$ in $\mathcal{D}_n$
                \STATE Sample mini-batch $\mathcal{B}_n$ from $\mathcal{D}_n$
                \FORALL{$(s,a,\tilde{r},s') \in \mathcal{B}_n$}
                    \STATE Compute target:
                    \[
                        y = \tilde{r} + \gamma V_{\psi_n}(s')
                    \]
                    \STATE Compute TD error:
                    \[
                        \delta = y - V_{\psi_n}(s)
                    \]
                    \STATE \textbf{Critic update:}
                    \[
                        \psi_n \leftarrow \psi_n + \eta_\psi \nabla_{\psi_n} \big( \delta^2 \big)
                    \]
                    \STATE \textbf{Actor (primal) update:}
                    \[
                        \theta_n \leftarrow \theta_n + \eta_\theta \, \nabla_{\theta_n} \log \pi_{\theta_n}(a \mid s)\, \delta
                    \]
                \ENDFOR
                \STATE \textbf{Dual (constraint) updates:}
                \FOR{$i = 1$ to $3$}
                    \STATE Estimate $\hat{g}_{i,n} \approx \mathbb{E}[g_{i,n}(t)]$ from recent samples
                    \STATE
                    \[
                        \lambda_{i,n} \leftarrow 
                        \big[ \lambda_{i,n} + \eta_\lambda\, \hat{g}_{i,n} \big]_+
                    \]
                \ENDFOR
            \ENDFOR
        \ENDFOR
    \end{algorithmic}
\end{algorithm}

\subsection{System Model}

We consider a dense terrestrial wireless network composed of set
\begin{equation}
    \mathcal{N} = \{1,2,\ldots,N\},
\end{equation}
of gNodeBs (gNBs), where gNB $n \in \mathcal{N}$ serves the dynamic set $\mathcal{U}_{n}(t)$ of user equipment (UEs) 
in time slot $t$. The network supports heterogeneous service slices for enhanced mobile broadband (eMBB), URLLC, and massive machine type communication (mMTC),
\begin{equation}
    \mathcal{S} = \{\text{eMBB},\ \text{URLLC},\ \text{mMTC}\},
\end{equation}
each associated with distinct QoS objectives. A traffic request from UE $u$ for slice $s$ at time $t$ is expressed as
\begin{equation}
    R_{u}^{s}(t) = \big\{ d_{u}^{s}(t),\ \lambda_{u}^{s}(t),\ \tau_{u}^{s}(t) \big\},
\end{equation}
where $d_{u}^{s}(t)$ is the required rate, $\lambda_{u}^{s}(t)$ is the traffic arrival rate, and $\tau_{u}^{s}(t)$ is the maximum tolerable latency. URLLC traffic possesses stringent reliability and delay requirements, and these constraints explicitly appear in the CMDP formulation.

\subsection{Channel, SINR Model, and Resource Allocation}

The wireless channel between gNB $n$ and UE $u$ at time $t$ is modeled as
\begin{equation}
    h_{n,u}(t) = \sqrt{\beta_{n,u}(t)}\, g_{n,u}(t),
\end{equation}
where $\beta_{n,u}(t)$ represents large-scale fading (path loss and shadowing), and $g_{n,u}(t)\sim\mathcal{CN}(0,1)$ models small-scale fading. ($h_{n,u}(t)$ is not explicitly used in the rest of this paper. If I understand correctly, you will use $H_n(t)$ the local CSI as the channel state. It would be better to state this. Also Eq. (11) also uses the symbol $\beta$, please use a different symbol here or Eq. (11). The resulting signal-to-interference-plus-noise Ratio (SINR) is
\begin{equation}
    \gamma_{n,u}(t) =
    \frac{P_n |h_{n,u}(t)|^2}
    {\sigma^2 + \sum_{\substack{m\in\mathcal{N}\\ m\neq n}}
    P_m |h_{m,u}(t)|^2},
\end{equation}
where $P_n$ is the gNB transmit power and $\sigma^2$ is the noise power. The achievable data rate is
\begin{equation}
    R_{n,u}(t) = B_{n,u}(t)\, \log_2\!\left(1+\gamma_{n,u}(t)\right),
\end{equation}
where $B_{n,u}(t)$ is the allocated bandwidth.

The resource allocation at each gNB is expressed as a continuous vector
\begin{equation}
    a_n(t) = \{ a_n^s(t) : s \in \mathcal{S} \},
\end{equation}
representing the physical resource block (PRB) fraction assigned to each slice, where 
\begin{equation}
    \sum_{s\in\mathcal{S}} a_n^s(t) \le 1,\qquad \forall n\in\mathcal{N}.
\end{equation}
\noindent\textit{Remark on Continuous Resource Allocation:}
Although PRBs are inherently discrete, we model the slicing decision $a_n(t)$ as a continuous-valued vector representing the fraction of available bandwidth assigned to each slice. This relaxation is standard in learning-based RAN slicing and is justified by the time-averaged nature of slice-level control, where decisions are applied over multiple scheduling intervals. Modeling $a_n(t)$ in a continuous simplex enables differentiable policy optimization, improves convergence stability in the constrained multi-agent setting, and allows the use of actor--critic and primal--dual learning methods. Discrete PRB assignments are subsequently realized by the lower-layer scheduler based on the learned slice-level proportions.

\subsection{Problem Formulation}

Each gNB is modeled as an autonomous RL agent operating a CMDP. The local state observed by gNB $n$ at time $t$ is
\begin{equation}
    s_n(t) = \big[ H_n(t),\ Q_n(t),\ a_n(t\!-\!1),\ D_n(t) \big],
\end{equation}
where $H_n(t)$ denotes local CSI, $Q_n(t)$ the per-slice queue lengths, $a_n(t-1)$ the previous allocation vector, and $D_n(t)$ a set of slice-performance indicators (e.g., throughput or latency violations).

The action taken by gNB $n$ is the resource allocation vector
\begin{equation}
    a_n(t) = \pi_{\theta_n}(s_n(t)),
\end{equation}
where $\pi_{\theta_n}$ is a parameterized decision policy.

Reward
\begin{equation}
    r_n(t) = 
    \sum_{s\in\mathcal{S}}
    \left(
    \omega_s \eta_n^s(t)
    - \alpha_s \delta_n^s(t)
    - \beta_s \rho_n^s(t)
    \right),
\end{equation}
captures system utility, QoS satisfaction, and control stability, where $\eta_n^s(t)= \sum_{u \in \mathcal{U}_n^s(t)} R_{n,u}(t)$ is the achieved throughput, $\delta_n^s(t)$ is the QoS violation penalty, and $\rho_n^s(t)=|a_n^s(t)-a_n^s(t-1)|$ is the reconfiguration cost (penalizing frequent or large changes in allocation).

Each gNB must additionally satisfy the following CMDP constraints:


We define the aggregate outgoing interference leakage generated by gNB $n$ at time $t$ as
\begin{equation}
    I_n(t)
    =
    \sum_{\substack{m \in \mathcal{N} \\ m \neq n}}
    \sum_{u \in \mathcal{U}_m(t)}
    P_n \, \big| h_{n,u}(t) \big|^2 ,
\end{equation}
which captures the total interference power imposed by gNB $n$ on users served by neighboring cells.

Each gNB $n$ is subject to multiple heterogeneous operational constraints, which are indexed by
$i \in \{1,2,3\}$ to enable a compact CMDP formulation and Lagrangian-based optimization.
Specifically, $g_{i,n}(t)$ denotes the instantaneous violation of the $i$-th constraint at gNB $n$
and time $t$, where each constraint captures a distinct physical or QoS requirement.

The corresponding CMDP constraint is expressed as
\begin{equation}
g_{1,n}(t) = \max\!\bigl(0,\, I_n(t) - I_n^{\max} \bigr),
\label{eq:g1}
\end{equation}

where $I_n^{\max}$ denotes the maximum tolerable interference leakage budget for gNB $n$.
Accordingly, $g_{1,n}(t)$ represents the \emph{inter-cell interference constraint violation}
at gNB $n$, enforcing an aggressor-side interference limit that is consistent with the
inter-cell interference term appearing in the SINR denominator of neighboring users.
The index $i=1$ simply identifies this constraint within the CMDP and does not imply any
ordering or priority.



Similarly, $g_{2,n}(t)$ and $g_{3,n}(t)$ denote the URLLC latency violation constraint
and the resource feasibility constraint, respectively. Collectively, the constraint set
$\{g_{i,n}(t)\}_{i=1}^3$ enables explicit enforcement of interference, latency, and
resource allocation limits within the CMDP framework.




The URLLC latency constraint
\begin{equation}
    g_{2,n}(t) =
    \sum_{u\in\mathcal{U}_n^{\text{URLLC}}(t)}
    \mathbb{I}\!\left\{\text{Delay}_u^s(t) > \tau_u^s(t)\right\},
\end{equation}
ensures reliability of latency-critical services.

The resource feasibility constraint is
\begin{equation}
    g_{3,n}(t) = \max\!\left(0,\ \sum_{s\in\mathcal{S}}a_n^s(t)-1\right).
    \label{eq:g3}
\end{equation}

The CMDP for gNB $n$ is defined as
\begin{equation}
    \mathcal{M}_n=\big(\mathcal{S},\mathcal{A},P,r_n,g_{1,n},g_{2,n},g_{3,n},\gamma\big).
\end{equation}
The network-wide goal is then
\begin{equation}
    \max_{\{\pi_{\theta_n}\}}
    \sum_{n=1}^N
    \mathbb{E}\!\left[
    \sum_{t=0}^{\infty} \gamma^t r_n(t)
    \right],
    \label{eq:cmdp_objective}
\end{equation}
subject to $\mathbb{E}[g_{i,n}(t)] \le 0$ for $i\in\{1,2,3\}$ and all $n$. 

\section{Lagrangian Relaxation for CRL}

\begin{algorithm}[t]
    \caption{Federated Constrained Multi-Agent RL for Dynamic Spectrum Slicing}
    \label{alg:federated_rl}
    \begin{algorithmic}[1]
        \STATE \textbf{Input:} Set of gNBs $\mathcal{N}$, initial global policy parameters $\theta^{(0)}$, number of federation rounds $K$, local update horizon $T_{\text{loc}}$, aggregation weights $\{w_n\}_{n\in\mathcal{N}}$
        \STATE Initialize global policy $\theta^{(0)}$ and broadcast to all gNBs
        \FOR{round $k = 0,1,\ldots,K-1$}
            \STATE Server selects participating subset $\mathcal{N}_k \subseteq \mathcal{N}$
            \FORALL{$n \in \mathcal{N}_k$ \textbf{in parallel}}
                \STATE Receive global policy $\theta^{(k)}$ from server
                \STATE Set local policy parameters $\theta_n \leftarrow \theta^{(k)}$
                \STATE Initialize/maintain local critic $\psi_n$ and dual variables $\lambda_n$
                \STATE Run \textbf{Algorithm~\ref{alg:local_primal_dual}} for $T_{\text{loc}}$ steps, obtaining updated $\theta_n^{(k)}$ and $\lambda_n^{(k)}$
                \STATE Upload $\theta_n^{(k)}$ (and optionally statistics of $\lambda_n^{(k)}$) to server
            \ENDFOR
            \STATE \textbf{Server-side aggregation:}
            \STATE Collect local parameters $\{\theta_n^{(k)} : n \in \mathcal{N}_k\}$
            \STATE Compute federated update (e.g., FedAvg):
            \begin{equation*}
                \theta^{(k+1)} = \sum_{n \in \mathcal{N}_k} w_n \theta_n^{(k)}, 
                \quad \sum_{n \in \mathcal{N}_k} w_n = 1
            \end{equation*}
            \STATE Optionally perform spectrum-aware policy distillation or regularization on $\theta^{(k+1)}$
            \STATE Broadcast updated global policy $\theta^{(k+1)}$ to all gNBs
        \ENDFOR
        \STATE \textbf{Output:} Trained global policy $\theta^{(K)}$ and converged local policies $\{\theta_n^{(K)}\}_{n\in\mathcal{N}}$
    \end{algorithmic}
\end{algorithm}

Directly solving the constrained optimization problem in
\eqref{eq:cmdp_objective} is intractable in high-dimensional and distributed
environments due to the continuous state--action spaces, coupled inter-cell
dynamics, and long-term constraint requirements. We therefore adopt a
Lagrangian-based primal--dual reinforcement learning framework to transform
the CMDP into an unconstrained saddle-point problem that can be solved
efficiently using policy gradient methods.
Let
\begin{equation}
    \lambda_n = [\lambda_{1,n},\lambda_{2,n},\lambda_{3,n}]^\top \ge 0
\end{equation}

denote the dual variables associated with the CMDP constraints defined in
\eqref{eq:g1}--\eqref{eq:g3}, namely,
the inter-cell interference constraint, the URLLC latency constraint, and the
resource feasibility constraint, respectively. Each dual variable $\lambda_{i,n}$ penalizes long-term violations of its
corresponding constraint in expectation.

\begin{equation}
    \mathcal{L}_n(\theta_n,\lambda_n) =
    \mathbb{E}\!\left[
    \sum_{t=0}^{\infty}
    \gamma^t
    \left(
    r_n(t) -
    \sum_{i=1}^3 \lambda_{i,n} g_{i,n}(t)
    \right)
    \right].
    \label{eq:Lagrangian}
\end{equation}
The expectation in \eqref{eq:Lagrangian} is taken over the state--action trajectories induced by policy $\pi_{\theta_n}$, making the dependence of $\mathcal{L}_n$ on $\theta_n$ implicit.
The constrained optimization problem becomes the saddle-point formulation
\begin{equation}
    \min_{\lambda_n \ge 0}\;
    \max_{\theta_n}\;
    \mathcal{L}_n(\theta_n,\lambda_n).
\end{equation}

The primal (policy) update is 
\begin{equation}
    \theta_n \leftarrow 
    \theta_n +
    \eta_\theta
    \nabla_{\theta_n}
    \mathcal{L}_n(\theta_n,\lambda_n)
\end{equation}
and the dual (constraint) update is
\begin{equation}
    \lambda_{i,n} \leftarrow
    \Big[
    \lambda_{i,n}
    + \eta_\lambda\ \mathbb{E}[g_{i,n}(t)]
    \Big]_+,
\qquad i=1,2,3.
\end{equation}
This primal--dual constrained RL formulation enables each gNB to autonomously explore feasible allocation strategies, enforce interference and QoS guarantees, and converge toward near-optimal slicing decisions. 

The complete local learning procedure executed at each gNB,
including the primal (policy) update, critic update, and dual
variable adaptation, is summarized in \textbf{Algorithm~1}.
Algorithm~1 details how each agent interacts with its local
environment, computes the Lagrangian-adjusted reward, and
updates its policy parameters while enforcing CMDP constraints.

\section{SliceFed}
\label{sec:proposed_method}
While Algorithm~1 describes the local constrained learning process
at each gNB, the overall SliceFed framework additionally incorporates
a federated coordination mechanism to enable collaborative training
across cells without sharing raw data. The global federated learning
workflow, including local model upload, aggregation, and redistribution,
is presented in \textbf{Algorithm~2}.

{SliceFed} is framework founded on the federated multi-agent DRL (F-MADRL) paradigm, where autonomous agents deployed at individual gNBs learn and execute local slicing policies in real time. SliceFed operates under strict resource, interference, and latency constraints, while a federated learning layer enables collaborative training of a globally coordinated policy without sharing sensitive user data.

We adopt the proximal policy optimization (PPO) as the underlying learning algorithm for local agents to effectively address the continuous nature of spectrum slicing decisions, the presence of time-varying inter-cell coupling, and the critical QoS requirements of heterogeneous services. PPO provides stable learning in continuous action spaces and is particularly robust under the non-stationary dynamics induced by multi-agent interactions in dense RAN environments. Its clipped policy updates prevent abrupt changes in the resource allocation that could otherwise lead to severe constraint violations, making it well suited for Lagrangian-based CRL.

Moreover, as an on-policy method, PPO naturally adapts to time-varying dual variables associated with CMDP constraints and to evolving interference conditions across neighboring cells. This property is critical in the federated multi-agent setting of SliceFed, where both the environment and constraint landscape change as agents learn concurrently. Building on this foundation, we detail the key components of the proposed architecture, including the federated multi-agent structure, the local constrained learning formulation, the collaborative training process, and the overall algorithmic workflow.

\subsection{Federated Multi-Agent Architecture}

We deploy a collection of autonomous slicing agents $\{ \mathcal{G}_n \}_{n=1}^N$, 
where agent $\mathcal{G}_n$ resides at gNB $n \in \mathcal{N}$.
Each agent learns its own local policy $\pi_{\theta_n}(a_n|s_n)$ by interacting with its local environment (i.e., its connected UEs, channel conditions, and queue states). The agents update their local model parameters $\theta_n$ independently based on these interactions.

After a set number of local training episodes, the agents’ learned knowledge
is aggregated to update the global model. At the beginning of global round
$k$, all agents initialize their local models with the global parameters
$\theta^{(k)}$. After completing local training, each agent obtains updated
parameters $\theta_n^{(k+1)}$, which are then aggregated by the central server
to produce the new global model $\theta^{(k+1)}$. We employ the Federated
Averaging (FedAvg) algorithm for this aggregation step. Specifically, in each
global communication round $k$, the central aggregator updates the global
model parameters as
\begin{equation}
\theta^{(k+1)} =
\sum_{n=1}^{N}
\frac{|\mathcal{D}_n|}{\sum_{j=1}^{N} |\mathcal{D}_j|}
\cdot \theta_n^{(k+1)} .
\end{equation}

where $|\mathcal{D}_n|$ is the size of the local dataset (i.e., the number of collected experience tuples) at gNB $n$ during round $k$. This weighted averaging ensures that agents contributing more experience have a proportionally larger impact on the global model.

\subsection{Local Agent: DRL Formulation}

Each local agent $\mathcal{A}_n$ formulates its decision-making problem as the local CMDP defined in Section~\ref{sec:system_model}.C. Eq. \eqref{eq:cmdp_objective}. An actor-critic DRL architecture is an effective choice for solving this, where an actor network learns the policy $\pi_{\theta_n}$ and a critic network estimates the value function to guide the actor's updates. The objective for each agent is maximizing its expected long-term cumulative return:
\begin{equation}
    \max_{\theta_n} \mathbb{E}_{s_n, a_n \sim \pi_{\theta_n}} \left[ \sum_{t=0}^{\infty} \gamma^t r_n(t) \right].
\end{equation}

The use of a local DRL agent allows each gNB to react swiftly to changes in its immediate environment, such as sudden traffic bursts from eMBB users or the arrival of a URLLC service request, without needing to communicate with a central controller for every decision.

\subsection{Collaborative Training and Synchronization}

Agents do not transmit raw local state information or UE data to maintain data privacy and minimize 
communication overhead. Instead, they only transmit their model parameter updates, $\Delta\theta_n = \theta_n^{(k+1)} - \theta_n^{(k)}$, to the central aggregator. This process significantly reduces the amount of data exchanged over the network.

Synchronization is not performed on a fixed schedule but is triggered dynamically. 
An aggregation round is initiated when the average training loss across the agents exceeds a predefined threshold: 
\begin{equation}
    \frac{1}{N} \sum_{n=1}^{N} L_n > \delta,
\end{equation}
where $L_n$ is the local loss of agent $n$'s DRL model. This ensures that communication resources are used only when the local models have sufficiently diverged from the global consensus, balancing communication efficiency with global model consistency.

\subsection{Policy Cohesion and Exploration}

In a dense network, the actions of one gNB can impact its neighbors through inter-cell interference. We introduce periodic policy distillation to ensure that the distributed policies remain coherent and stable. This involves adding a loss term that minimizes the divergence between an agent's local policy $\pi_{\theta_n}$ and the global policy $\pi_{\theta}$:
\begin{equation}
    \mathcal{L}_{\text{distill}} = \sum_{n=1}^{N} ||\pi_{\theta_n}(s) - \pi_{\theta}(s)||^2.
\end{equation}
This encourages local policies to stay close to the globally learned strategy, preventing catastrophic interference or resource starvation. Furthermore, we add an entropy regularization term to the actor's loss function,
\begin{equation}
    \mathcal{L}_{\text{total}} = \mathcal{L}_{\text{actor}} + \lambda_H \cdot \mathcal{H}[\pi_{\theta_n}],
\end{equation}
where $\mathcal{H}[\pi_{\theta_n}]$ is the entropy of the policy and $\lambda_H$ is a weighting coefficient. This encourage agents to explore new and potentially better slicing strategies.

\subsection{The SliceFed Algorithm}
Algorithm~\ref{alg:slicefed} describes the complete training and inference procedure for the SliceFed framework. The process alternates between parallel local model training at each gNB and synchronous global model aggregation at a central server.


\begin{algorithm}[h!]
\caption{The SliceFed Framework for Dynamic RAN Slicing}
\label{alg:slicefed}
\begin{algorithmic}[1]
\STATE \textbf{Initialize:} Global model $ \theta^0 $; local agents $ \mathcal{A}_n $ at each gNB with $ \theta_n^0 = \theta^0 $.
\FOR{each global round $ k = 0, 1, \dots $}
    \FOR{each agent $ n = 1, \dots, N $ at each gNB in parallel}
        \STATE Collect local trajectory $ \tau_n^k = \{(s_n, a_n, r_n)_t\} $ over several time steps.
        \STATE Update local policy: $ \theta_n^{k+1} \leftarrow \text{Update}(\theta_n^k, \tau_n^k) $ using a DRL algorithm.
        \STATE Send model update $ \Delta\theta_n = \theta_n^{k+1} - \theta_n^k $ to central aggregator.
    \ENDFOR
    \STATE \textbf{Aggregate global model:}
    \[
        \theta^{k+1} \leftarrow \text{FedAvg}(\{\theta_n^{k+1}\}) \quad \text{using Eq. (10)}
    \]
    \STATE \textbf{Broadcast} updated global model $ \theta^{k+1} $ to all gNB agents.
\ENDFOR
\end{algorithmic}
\end{algorithm}


\section{Numerical Analysis}
\label{sec:simulation_results}

We evaluate the performance of the proposed SliceFed framework through rigorous simulations in a dense multi-cell RAN environment. We assess how 
F-MARL improves spectrum slicing efficiency, QoS satisfaction, and robustness compared to conventional and learning-based baselines.

\subsection{Simulation Setup}

We consider a dense RAN composed of $N=7$ regularly 
deployed gNBs, each serving $U=10$ users per cell within a system bandwidth of $B=20$~MHz. The network supports three heterogeneous service slices: eMBB, URLLC, mMTC. Time is slotted with a scheduling interval of $1$~ms.

Wireless channels are modeled using a composite fading environment, incorporating distance-based path loss ($\alpha=3.7$), log-normal shadowing ($\sigma=6$~dB), and Rayleigh small-scale fading. We explicitly model inter-cell interference by calculating the aggregate signal leakage from simultaneous transmissions across neighboring gNBs. The maximum interference budget is set to $I_{max} = -15.0$~dBm.

Traffic arrivals for each slice follow independent Poisson processes. Unless otherwise specified, the default traffic loads are set to $\lambda_{eMBB}=1.5$, $\lambda_{URLLC}=4.0$, and $\lambda_{mMTC}=1.0$ packets/slot. URLLC packets are subject to a hard latency deadline of $\tau = 1$~ms.

\subsubsection{Hyperparameters}
Each gNB operates an autonomous PPO based actor-critic agent implemented in PyTorch. The actor and critic networks comprise two hidden layers with 128 neurons each with the ReLU activation function. We use the Adam optimizer with a learning rate of $3 \times 10^{-3}$ for both networks. Constraint satisfaction is enforced via a Lagrangian primal-dual mechanism with dual learning rates $\eta_{\lambda} = [0.01, 0.01, 0.01]$. The federated learning process runs for $200$ global communication rounds, with local updates performed over trajectories of $1000$ time steps. 
Results are averaged over $5$ random seeds.

\subsection{Baselines}

We compare SliceFed against three representative baselines covering static, heuristic, and stochastic strategies:
\begin{itemize}
    \item \textbf{Equal Slicing:} Resources are equally divided among slices 
    at each gNB, regardless of the traffic state.
    \item \textbf{Queue-Proportional (QueueProp):} Bandwidth is allocated proportional to the instantaneous queue lengths of each slice.
    \item \textbf{Random Allocation:} Resource fractions are sampled from a symmetric Dirichlet distribution at every time step.
\end{itemize}

\subsection{Learning Convergence}

\begin{figure}[t]
    \centering
    \includegraphics[width=\linewidth]{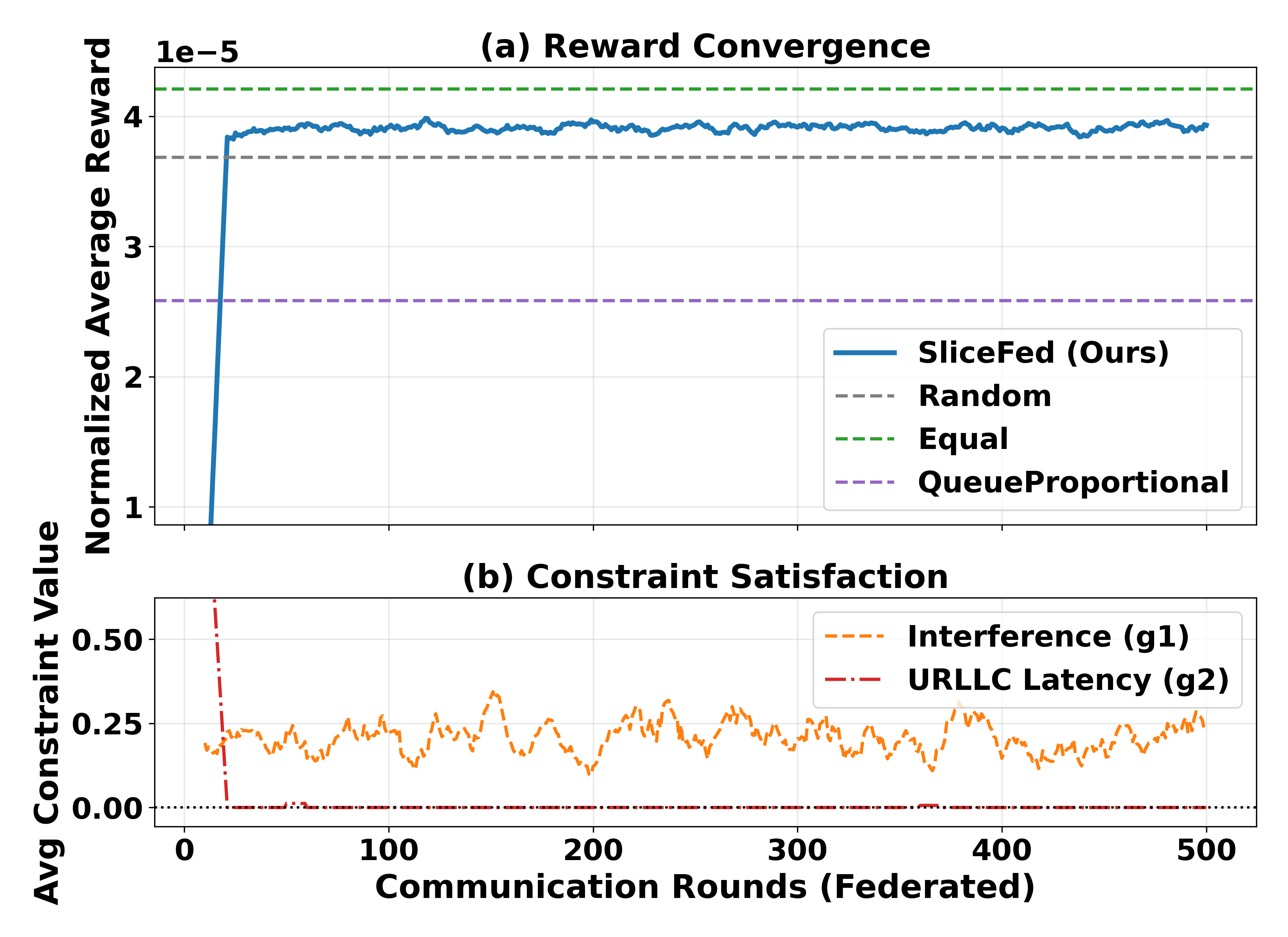}
    \caption{Training dynamics of SliceFed over federated communication rounds: Normalized average reward compared to baselines(a) and evolution of constraint violations for Interference ($g_1$) and URLLC Latency ($g_2$) (b).}
    \label{fig:fig1}
\end{figure}

Fig.~\ref{fig:fig1} illustrates the convergence behavior of SliceFed. As shown in Fig.~\ref{fig:fig1}(a), the federated PPO agents converge rapidly, stabilizing around a normalized reward of $3.9 \times 10^{-5}$ within 50 communication rounds. While the static Equal Slicing baseline achieves a slightly higher raw reward, it does so by over-provisioning resources without dynamic interference awareness. SliceFed learns a policy that trades off a marginal amount of throughput to
strictly satisfy the imposed CMDP safety constraints, including inter-cell interference limits, URLLC latency requirements, and resource feasibility.


Fig.~\ref{fig:fig1}(b) demonstrates the convergence of the proposed primal--dual
learning mechanism. The URLLC latency constraint ($g_2$, red curve) shows
temporary violations during early exploration but is driven to zero as the dual variables adapt, indicating that the latency constraint is ultimately satisfied. Concurrently, inter-cell interference ($g_1$, orange curve) is regulated near the feasibility boundary defined by the maximum allowable interference leakage $I_n^{\max}$. The learned policy maintains $I_n(t)$ close to this budget, maximizing spectral efficiency while avoiding constraint violations, which reflects the intended trade-off enforced by the Lagrangian updates.


\subsection{URLLC Reliability Analysis}

\begin{figure}[t]
    \centering
    \includegraphics[width=\linewidth]{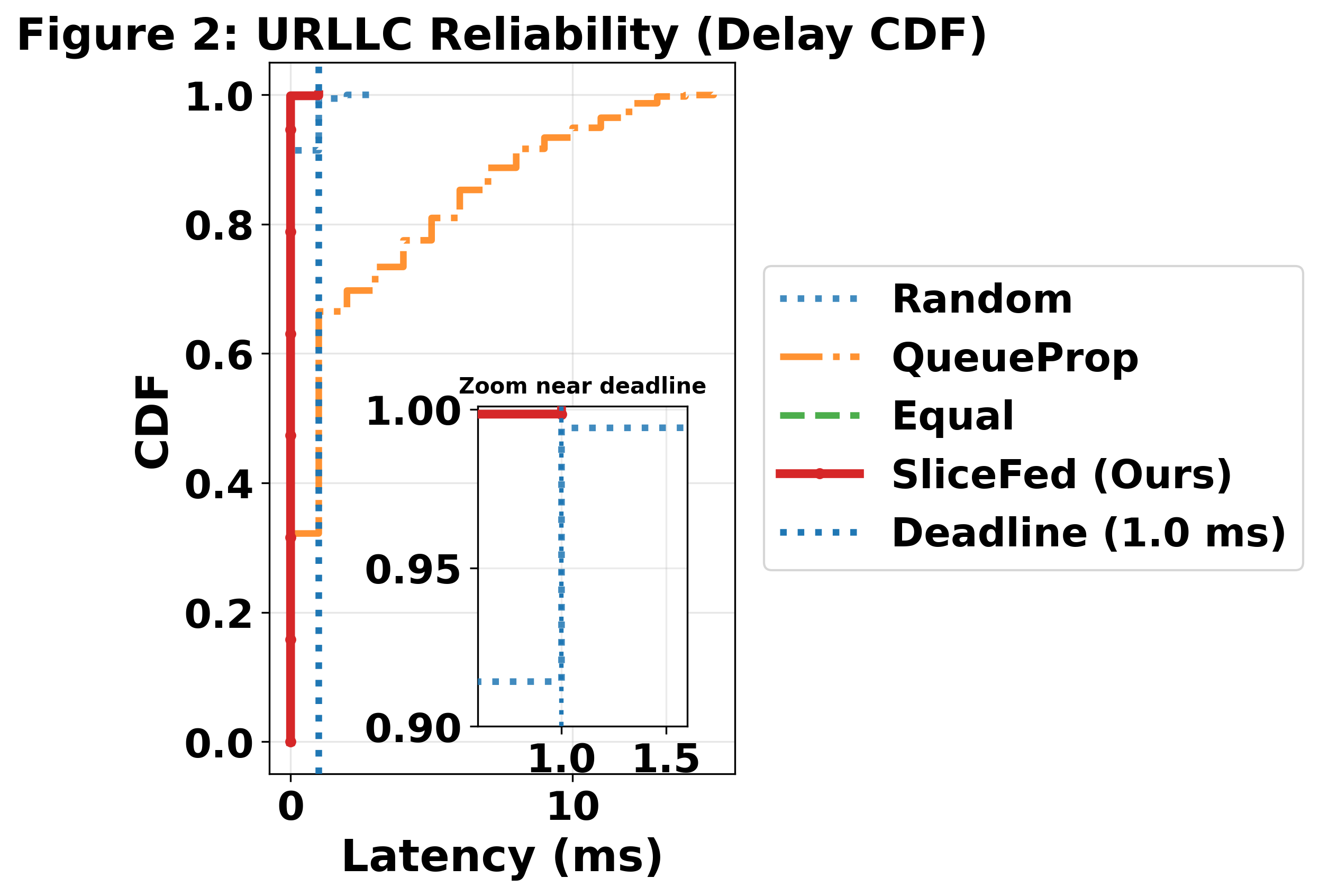}
    \caption{Empirical CDF of URLLC packet delays. 
    }
    \label{fig:fig2}
\end{figure}

Fig.~\ref{fig:fig2} plots the empirical cumulative distribution function (CDF) of URLLC packet delays. 
We make the following observations:
\begin{itemize}
    \item {SliceFed (blue)} demonstrates superior reliability, with the CDF reaching $\approx 1.0$ at the deadline. This confirms that the learned policy effectively prioritizes URLLC traffic to satisfy the $g_2$ constraint.
    \item {QueueProp (green)} exhibits a heavy-tailed distribution, failing to serve approximately 40\% of packets by the deadline. This highlights the failure of reactive, queue-based heuristics which cannot anticipate the stringent timing requirements of URLLC.
    \item {Equal Slicing (orange)} achieves high reliability by allocating 33\% of resources to URLLC regardless of load, which is spectrally inefficient compared to SliceFed's adaptive approach.
\end{itemize}

\subsection{Queue Dynamics and Resource Allocation}

\begin{figure}[t]
    \centering
    \includegraphics[width=\linewidth]{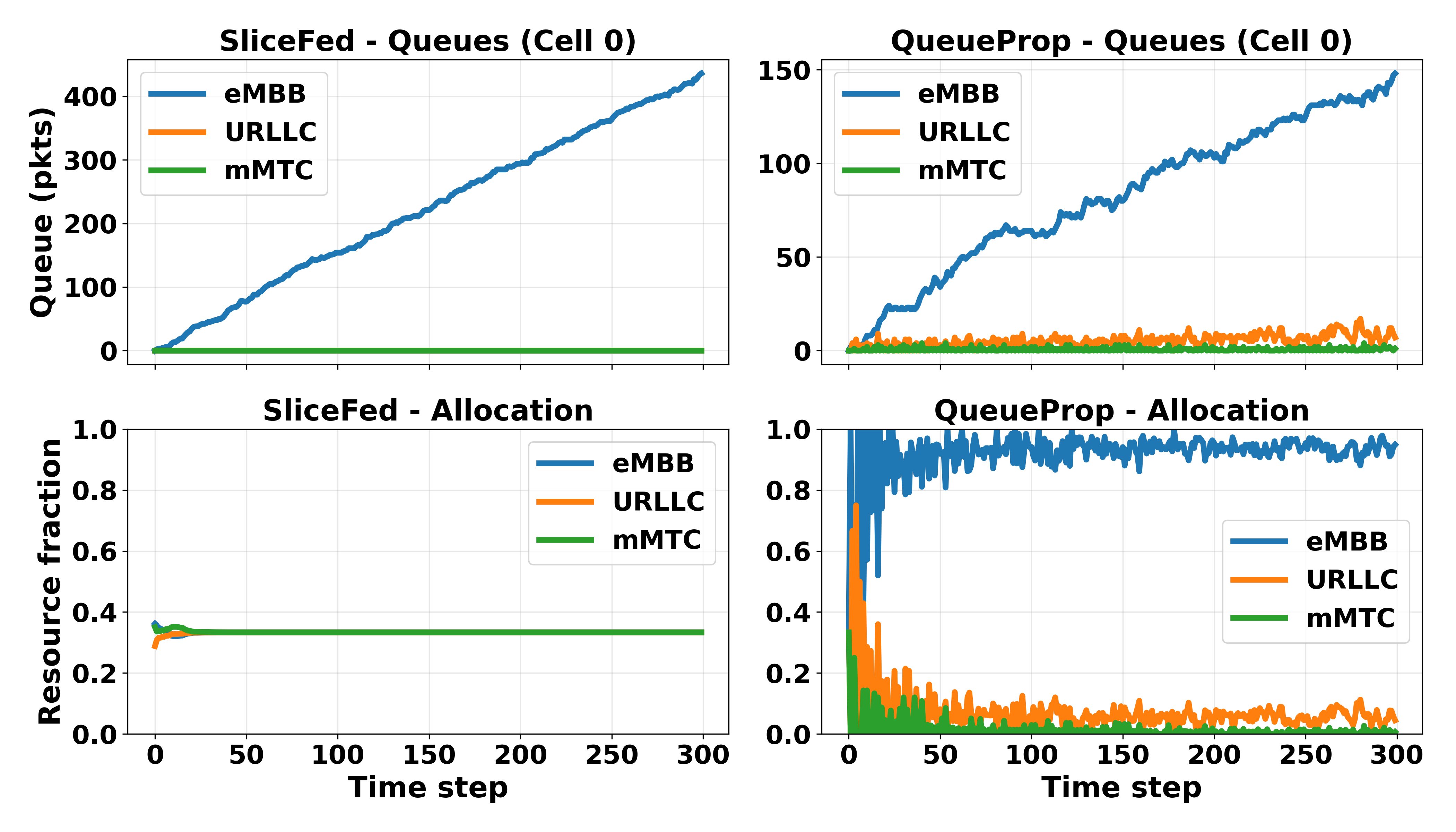}
    \caption{Temporal evolution of queues (top) and resource allocation (bottom) for SliceFed (left) and QueueProp (right).
    }
    \label{fig:fig3}
\end{figure}

Fig.~\ref{fig:fig3} contrasts the temporal dynamics of SliceFed and QueueProp. Under SliceFed, the distributed SliceFed agents learn to keep URLLC and mMTC queues near zero. The eMBB queue grows linearly, reflecting the system's saturation state. SliceFed correctly identifies eMBB as delay-tolerant, deprioritizing it to protect critical services. The SliceFed resource allocation converges to a highly stable operating point with minimal variance. This stability prevents the ping-pong effects and signaling overhead associated with rapid reconfiguration.
In contrast, QueueProp suffers from oscillatory behavior. Because it reacts solely to queue lengths, it allows URLLC queues to build up before allocating resources, resulting in jitter and unreliable latency performance.

\subsection{Robustness to Traffic Load}

\begin{figure}[t]
    \centering
    \includegraphics[width=\linewidth]{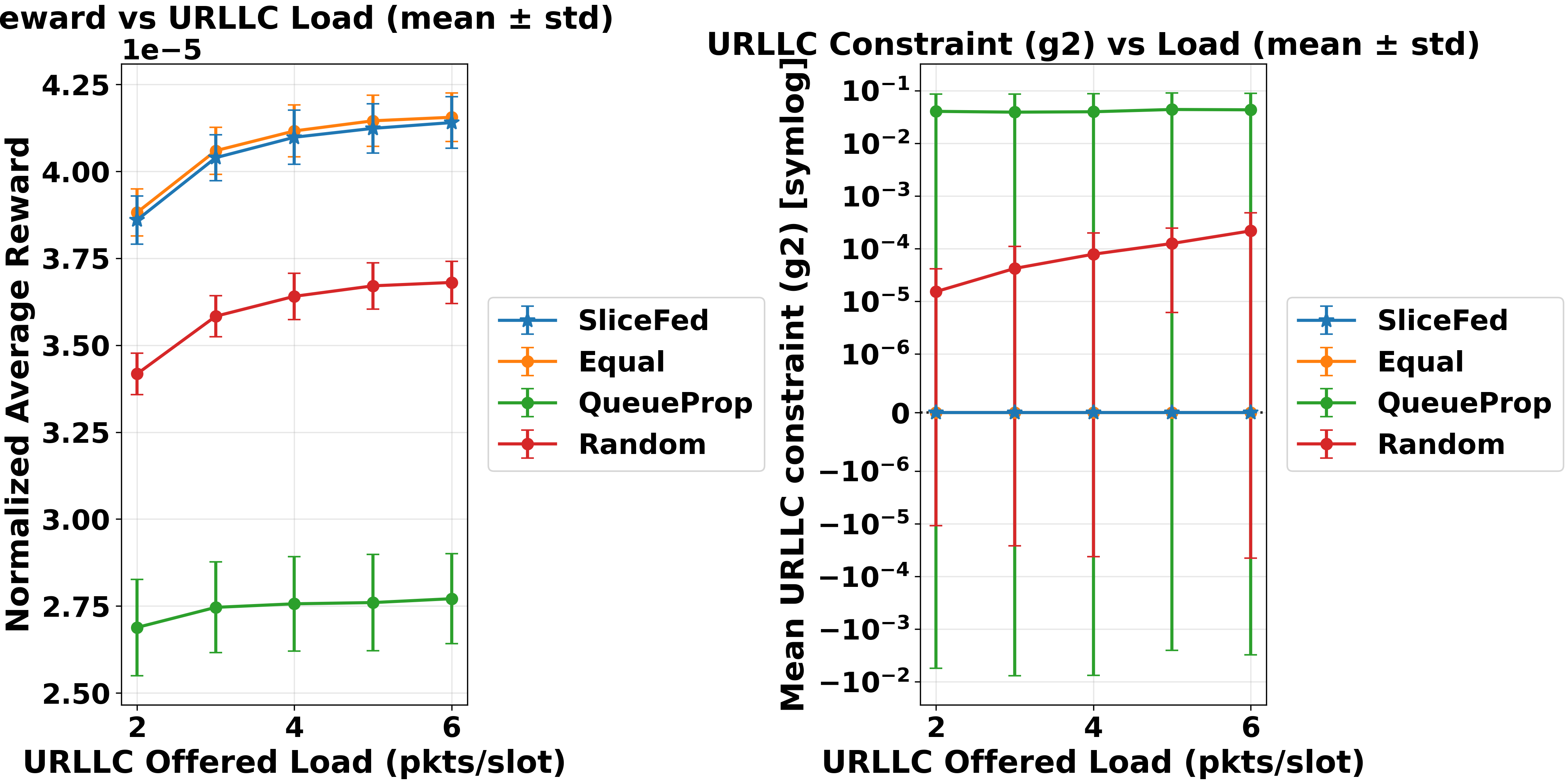}
    \caption{
    Robustness analysis under increasing URLLC traffic load~$\lambda$.
    Left: average normalized reward (mean~$\pm$~std).
    Right: mean URLLC latency constraint value~$g_2$ (mean~$\pm$~std).
    SliceFed maintains $g_2 \approx 0$ across all traffic loads, indicating strict
    satisfaction of the URLLC latency constraint. The apparent negative values
    visible for some baselines arise from statistical averaging and standard
    deviation visualization around zero on a symmetric log scale and do not
    correspond to negative constraint violations.
    }
    \label{fig:fig4}
\end{figure}

Fig.~\ref{fig:fig4} evaluates the robustness of the trained policy by sweeping the URLLC offered load from 2 to 6 packets/slot.
While SliceFed achieves a lower raw reward than the Equal Slicing and Random Allocation baselines (Fig.~\ref{fig:fig4}a), this metric must be interpreted in the context of safety. The baselines achieve higher throughput 
by ignoring the constraints. 
SliceFed maintains a constraint violation of effectively zero ($g_2 \approx 0$) across the entire load regime. In contrast, Random Allocation and QueueProp exhibit significant violations, 
rendering them unusable for mission-critical 6G applications (Fig.~\ref{fig:fig4}b). 

\subsection{Summary of Findings}
Across all experiments, SliceFed consistently outperforms the baseline methods by
\begin{enumerate}
    \item Enforcing strict hard QoS guarantees for URLLC traffic through explicit Lagrangian constraint learning,
    \item Achieving stable, low-variance resource allocation policies that reduce control overhead and reconfiguration oscillations, and
    \item Maintaining superior performance and constraint satisfaction across a wide range of URLLC traffic loads, demonstrating strong robustness and generalization beyond the training operating point.
\end{enumerate}
Unlike SliceFed, baseline methods exhibit traffic-insensitive behavior due to static allocation (Equal Slicing) or reactive queue-based control without explicit constraints (QueueProp), leading to persistent inefficiencies or violations as traffic conditions vary.

\section{Conclusions}

This paper has proposed SliceFed, a 
F-MARL framework for dynamic spectrum slicing in dense 6G RANs. By modeling the resource allocation problem as a CMDP, autonomous agents learn to maximize throughput while strictly adhering to inter-cell interference budgets and hard URLLC latency constraints. Simulation results confirm that SliceFed significantly outperforms heuristic and unconstrained baselines, achieving superior reliability, allocation stability, and robustness to traffic surges. Future work will investigate asynchronous federated aggregation and integration with O-RAN.

\section*{Acknowledgment}
\footnotesize The work by Mohammadi and Marojevic was supported in part by NSF Award 
2332661. 

\bibliographystyle{ieeetr}
\bibliography{Bib/main-bib.bib}

\end{document}